\documentclass[conference]{IEEEtran}
\IEEEoverridecommandlockouts

\usepackage{cite}
\usepackage{amsmath,amssymb,amsfonts}
\usepackage{algorithmic}
\usepackage{graphicx}
\usepackage{textcomp}
\usepackage{xcolor}

\usepackage{booktabs}
\usepackage{tabularx}
\usepackage{multirow}
\usepackage{makecell}
\usepackage{colortbl}

\usepackage{pgfplots}
\pgfplotsset{compat=1.18}
\usepackage{placeins}

\usepackage{enumitem}
\usepackage{csquotes}
\usepackage{soul}
\usepackage{pifont}

\usepackage{framed}
\usepackage{tcolorbox}
\usepackage{fancybox}

\usepackage{url}
\usepackage[hidelinks,bookmarks=false]{hyperref}

\usepackage{fontawesome}
\usepackage{orcidlink}

\usepackage{calc}
\usepackage{multicol}

\usepackage{comment}

\usepackage{tikz}

\newcommand{\xmark}{\ding{55}}

\def\BibTeX{{\rm B\kern-.05em{\sc i\kern-.025em b}\kern-.08em
    T\kern-.1667em\lower.7ex\hbox{E}\kern-.125emX}}
\begin{document}

\makeatletter
\newcommand{\linebreakand}{%
  \end{@IEEEauthorhalign}
  \hfill\mbox{}\par
  \mbox{}\hfill\begin{@IEEEauthorhalign}
}
\makeatother

\newcommand\copyrighttext{%
  \footnotesize \textcopyright~2026 IEEE. This document is a preprint. Personal use of this material is permitted.
  Permission from IEEE must be obtained for all other uses, in any current or future
  media, including reprinting/republishing this material for advertising or promotional
  purposes, creating new collective works, for resale or redistribution to servers or
  lists, or reuse of any copyrighted component of this work in other works. 
}

\newcommand\copyrightnotice{%
  \begin{tikzpicture}[remember picture,overlay]
  \node[anchor=north,yshift=-12pt] at (current page.north) {\fbox{\parbox{\dimexpr\textwidth-\fboxsep-\fboxrule\relax}{\copyrighttext}}};
  \end{tikzpicture}%
}

\title{From Quality Properties to Practice: A Guideline and Workflow for Explainability Requirements}
\author{\IEEEauthorblockN{Anonymous Authors}}

\author{
    \IEEEauthorblockN{
        Martin Obaidi\orcidlink{0000-0001-9217-3934}, 
        Jakob Droste\orcidlink{0000-0001-8746-6329}, 
        Hannah Deters\orcidlink{0000-0001-9077-7486}, 
        Marc Herrmann\orcidlink{0000-0002-3951-3300},
        Michel Krahl, 
        Kurt Schneider\orcidlink{0000-0002-7456-8323}}
    \IEEEauthorblockA{\textit{Leibniz Universität Hannover} \\
    \textit{Software Engineering Group}\\
    Hannover, Germany \\
    \{martin.obaidi, jakob.droste, hannah.deters, marc.herrmann, kurt.schneider\}@inf.uni-hannover.de}
}

\maketitle

\copyrightnotice
\vspace{-2ex}

\begin{abstract}
Explainability is increasingly required in AI-enabled software systems to support transparency, user trust, and compliance. Yet, explainability requirements are often written ad hoc, and unguided large language model support can yield vague, inconsistent, or incomplete statements. This paper presents a sequential, guideline-driven workflow for formulating explainability requirements and evaluates its tool-based operationalization. We first elicited candidate quality properties through a structured literature review and developer interviews. We then prioritized these properties in an online survey with practitioners ($n=20$) and derived a concise guideline of ten core properties with actionable formulation instructions. Next, we operationalized the guideline in a web-based tool that supports an iterative workflow of drafting, property-based checks, and revision. We evaluated the workflow in two complementary studies. In a task-based study with requirements engineers ($n=6$), formulation time was 23.5\% lower with tool support (mixed-effects model $p=0.049$, Wilcoxon sensitivity analysis $p=0.021$). In an independent online study with software developers ($n=18$), tool-supported and manually written requirements did not differ significantly in implementability or formulation quality, with a descriptive slight preference tendency toward the tool-supported versions. Overall, our results suggest that combining a prioritized quality guideline with lightweight LLM support can reduce formulation effort without significant differences in perceived quality from manually written requirements.
\end{abstract}


\section{Introduction}
\label{sec:intro}

Software systems support our everyday lives in almost all areas, both professional and private~\cite{Koehler2013,shklovski2014,li2022}. At the same time, their functionality and socio-technical environments are growing in complexity. Therefore, users increasingly want to understand how a system behaves and why it produces certain outcomes~\cite{adadi2018peeking,anders2022mentalmodels}. When a system lacks transparency and understandability, users may develop an explanation need~\cite{droste2024explanations,droste2025operationaltaxonomy}. If an explanation need is not addressed, this may lead to reduced trust or even rejection of the system. For development teams, such signals create pressure to respond, not only by providing ad hoc answers, but also by improving the system so that similar issues can be addressed systematically.

Requirements engineering provides the natural mechanism for this systematic response. Explainability is often treated as a non-functional requirement, but in practice it is difficult to specify precisely. Explainability requirements must capture who needs an explanation, in which context, in what form, with which level of detail, and with which acceptance or verification criteria~\cite{brunotte2023context,chazette2021exploring}. If these requirements are underspecified, the resulting solutions are inconsistent or hard to validate. If they are overspecified, they can become infeasible or overly restrictive. In short, the formulation quality of explainability requirements strongly influences whether explainability features can be implemented, tested, and maintained effectively.

One possible way to support formulation tasks is to use large language models (LLMs). LLMs can draft requirements quickly, yet unguided generation often yields text that is vague, incomplete, or inconsistent with intended system behavior~\cite{obaidi2025explainreqs}. This is particularly problematic for explainability, where requirements must remain verifiable despite describing user-facing communication. While the creation and presentation of explanations has been examined in several works, including approaches informed by XAI algorithms in AI-enabled systems~\cite{mavrepis2024,stepin2021} and methods for generating and presenting explanations~\cite{lubos2024,kabir2024}, there is still a gap in specifying well-formulated explainability requirements that provide sufficiently concrete guidance for development. What is missing is not another unguided generation setup, but a practical, empirically grounded guideline that makes requirement quality properties actionable and a way to operationalize this guideline so that teams can apply it consistently.

This paper addresses this gap by moving from quality properties to an operational workflow for writing explainability requirements. In contrast to prior work that primarily supports eliciting, classifying, generating, or presenting explanations, our focus is the formulation step: turning an identified explanation need into a guideline-aligned requirement. We first identify candidate properties of well-formulated explainability requirements from research and practice. We then prioritize these properties through an online survey to derive a concise guideline with concrete formulation instructions. Next, we operationalize the guideline in a web-based formulation tool that supports an iterative workflow of generation, guideline-based quality checks, and revision. Finally, we evaluate the tool-supported workflow in two complementary studies. A task-based study with requirements engineers assesses efficiency by measuring formulation time, and an independent online study with software developers assesses perceived implementability and formulation quality of the resulting requirements.

The central contributions of this work are as follows:
\begin{itemize}
    \item We provide a prioritized formulation guideline for explainability requirements, grounded in a structured literature review, developer interviews, and an online prioritization survey.
    \item We construct an iterative workflow that operationalizes these properties as actionable formulation instructions for writing explainability requirements.
    \item We operationalize the workflow in a web-based LLM-supported tool and evaluate it with respect to formulation time, perceived implementability, and formulation quality.
    \item We provide the tool implementation and study materials in a replication package to support reuse and follow-up research~\cite{obaidi2026guidelineReqDataset}.
\end{itemize}

The rest of this paper is structured as follows. Section~\ref{sec:background} reviews relevant background and related work. Section~\ref{sec:research} presents the study design. Section~\ref{sec:guideline-derivation} reports how we derived the guideline and its results. Section~\ref{sec:tool-evaluation} presents the tool-based evaluation and its results. Section~\ref{sec:discussion} discusses the findings. Finally, Section~\ref{sec:conclusion} concludes the paper.

\section{Background and Related Work}
\label{sec:background}

\subsection{Software Explainability}
\label{sec:software-explainability}

Software explainability has attracted substantial attention in recent years, including in requirements engineering~\cite{chazette2020explainability,kohl2019explainability,chazette2021exploring,obaidi2025appfeatures,droste2026immersive,droste2026misunderstandings,obaidi2026guidelineExp,obaidi2026usefulness,deters2025explanationcatalog}. One key challenge is that explanation needs are highly individual and depend on both users and context~\cite{ramos2021modeling}. In addition, different stakeholder groups, such as engineers, end users, or legal professionals often require different kinds of explanations~\cite{kohl2019explainability}. This variability makes it essential to elicit explainability requirements carefully~\cite{droste2024explanations}.

To support structured reasoning about explanation needs, a number of taxonomies have been proposed~\cite{unterbusch2023explanation,droste2024explanations,speith2022XAITaxonomies}. Unterbusch et al.~\cite{unterbusch2023explanation} derived a taxonomy of explanation needs from app reviews and distinguished primary from secondary concerns. Droste et al.~\cite{droste2024explanations} developed a taxonomy for everyday software based on an online survey and found that the categories \textit{interaction} and \textit{system behavior} were most prevalent. Speith~\cite{speith2022XAITaxonomies} reviewed eleven existing explainability taxonomies and provided recommendations for their design and use. 

\subsection{Requirement Elicitation}

\subsubsection{Requirement Elicitation in General}

Requirement elicitation is a core activity in requirements engineering, as it determines the foundation for subsequent development work~\cite{pacheco2018requirements}. Involving relevant stakeholders is essential to ensure that the system reflects their needs and expectations~\cite{mishra2008successful}. Common elicitation methods include interviews, surveys, observations, focus groups, brainstorming, and prototyping~\cite{obaidi2025elicitation,younas2017non,alflen2020model,pacheco2018requirements}.

For projects characterized by complexity and frequent change, Mishra et al.~\cite{mishra2008successful} recommend combining interviews, workshops, and iterative development to improve accuracy and completeness. Hadar et al.~\cite{hadar2014role} studied the role of domain knowledge in interview-based elicitation and found that domain knowledge can both support and hinder communication and the understanding of stakeholder needs.

\subsubsection{Elicitation of Explainability Requirements}
\label{sec:elicitation-of-explainability-requirements}

Research on eliciting explainability requirements has expanded rapidly~\cite{droste2024explanations,unterbusch2023explanation,obaidi2025appKnowledge,obaidi2025mood,obaidi2025automatingexplanationneedmanagement,deters2024qualitymodel,Deters2025quality,obaidi2025elicitation}. Across these works, one recurring difficulty is that directly asking users which explanations they need may distort responses due to hypothetical bias~\cite{hypotheticalBiasPlott} and the \textit{why-not mentality}~\cite{droste2023designing}. The \textit{why-not mentality} describes users' tendency to agree that they want an explanation when asked, because potential downsides are not salient to them. To reduce these biases, Deters et al.~\cite{deters2024pulse} explored whether biometric data can indicate explanation needs. However, their results suggest that biometric signals are not yet reliable predictors, warranting further research.

Several approaches aim to structure or support requirements elicitation for explainable systems. Ramos et al.~\cite{ramos2021modeling} proposed explainability personas based on questionnaire responses from 61 users and derived five personas representing distinct explanation needs in recommender systems. Droste et al.~\cite{droste2024explanations} derived a taxonomy of explanation needs for everyday software. Chazette et al.~\cite{chazette2022framework} proposed a quality framework for analyzing and operationalizing explainability requirements and validated it in a navigation app case study.

Beyond identifying explanation needs, recent work has also compared elicitation methods and examined how taxonomies shape the elicitation outcome. Obaidi et al.~\cite{obaidi2025elicitation} studied focus groups, interviews, and surveys in an industrial case study and found that interviews were most efficient in terms of distinct needs per participant per time spent, while surveys collected the largest number of needs but with higher redundancy. They also reported that introducing a taxonomy after an initial free elicitation phase increased the number and diversity of needs, suggesting benefits of a two-phase approach that combines open elicitation with later structuring.

\subsection{Requirement Specification and LLM Support}
\label{sec:specification-and-llms}

While elicitation methods help gather user needs, translating these needs into well-formulated and verifiable specifications remains a manual and error-prone task. Recently, the requirements engineering community has increasingly explored the use of LLMs to support this formulation process. Arora et al.~\cite{Arora.2023} provide early insights into the potential of generative AI in requirements engineering, demonstrating that LLMs can fundamentally support tasks such as requirement elicitation, analysis, specification, and validation.

However, applying LLMs to RE tasks also introduces significant challenges. Norheim~\cite{Norheim.2024} describes key difficulties in this context, emphasizing the need for clear prompt structures to ensure consistency and traceability in the generated outputs. Without structured guidance, unguided LLM-based generation often yields requirements that are imprecise, inconsistent, or lack necessary verification criteria~\cite{obaidi2025explainreqs}. 

To mitigate these issues, targeted prompt engineering is one possible support mechanism, but it does not remove the need for human validation. Chen et al.~\cite{Chen.2025} demonstrate that prompt engineering significantly impacts the quality of LLM outputs, particularly for formal and structured texts such as requirements. 

\subsection{Positioning of This Work}

Prior work provides important foundations for identifying explanation needs, structuring them through taxonomies, and analyzing explainability as a quality concern. However, these works leave open how an identified explanation need can be turned into a consistently formulated requirement. Our work addresses this formulation step. It combines empirically prioritized requirement quality properties with an iterative workflow for drafting, checking, and revising explainability requirements. In contrast to unguided LLM generation, the LLM is used as lightweight support within a guideline-driven process rather than as an autonomous source of requirements.

\section{Study Design}
\label{sec:research}

\begin{figure*}[htbp]
    \centering
    \includegraphics[width=0.8\linewidth]{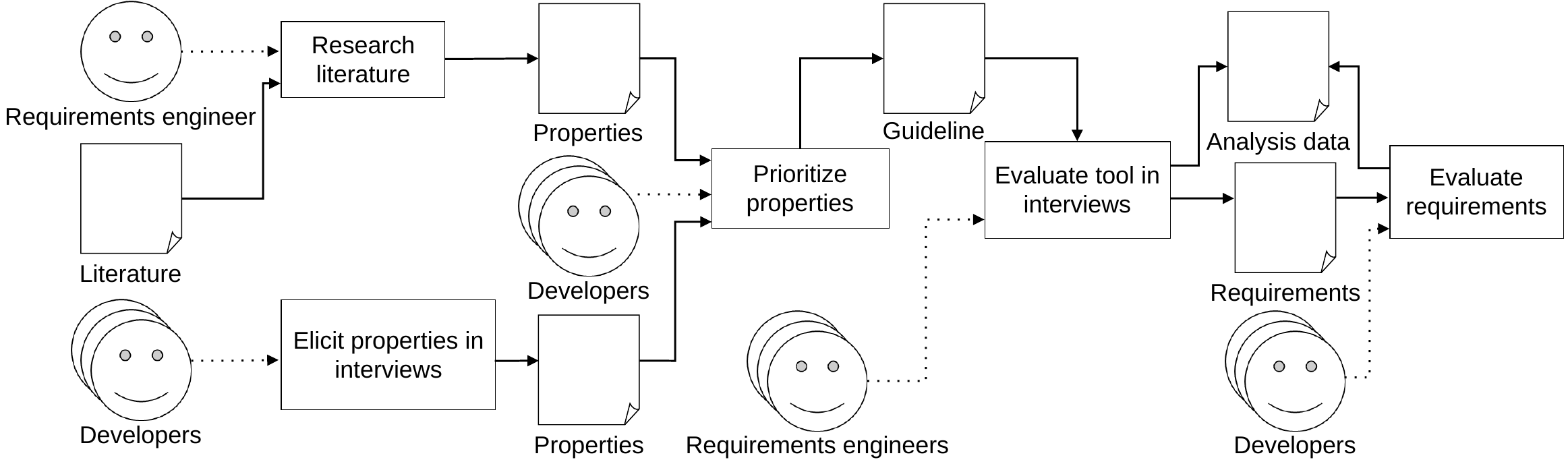}
    \caption{Overview of our research design in FLOW notation~\cite{stapel2009flow}.}
    \label{fig:overview-study}
\end{figure*}

\subsection{Research Goal}
\label{sec:research-goal}

Prior work on ad hoc and fully automated generation of explainability requirements indicates that unguided LLM-based approaches often yield requirements that are imprecise, inconsistent, and incomplete~\cite{obaidi2025explainreqs}. In this paper, we therefore investigate a sequential, guideline-driven approach that (i) derives quality properties for explainability requirements from research and practice, (ii) prioritizes these properties to form a concise guideline, and (iii) operationalizes the guideline in a web-based formulation tool that supports iterative generation, checking, and refinement.

We strive to achieve the following goal, formulated according to Wohlin et al.'s Goal-Definition Template~\cite{wohlin2012experimentation}:\\

\setlength{\shadowsize}{2pt}
\noindent
\shadowbox{
\begin{minipage}[t]{0.95\columnwidth}
\textbf{Research Goal:} \textit{Evaluate} a guideline-driven, LLM-supported formulation process and tool
\textit{for the purpose of} understanding its impact on supporting the formulation of well-formulated explainability requirements
\textit{with respect to} (i) the identification and prioritization of quality properties, and (ii) efficiency and perceived quality of resulting requirements compared to manual formulation
\textit{from the point of view of} requirements engineers and software developers
\textit{in the context of} specifying explainability requirements for AI-enabled software systems.
\end{minipage}
}\vspace{0pt}

\subsection{Research Questions}
\label{sec:research-questions}

To structure our sequential study—from eliciting candidate quality properties, to prioritizing them into a guideline, to operationalizing them in a tool and evaluating its effects—we address the following research questions:

\begin{itemize}
    \item \textbf{RQ1: Which properties characterize well-formulated explainability requirements from research and practice?}\\
    Answering this research question helps us establish a consolidated set of quality properties grounded in both literature and practitioner experience. This set serves as the foundation for constructing a guideline for formulating explainability requirements.

    \item \textbf{RQ2: Which of these properties are perceived by software developers as most \textit{useful} for formulating explainability requirements?}\\
    This question allows us to prioritize the elicited properties based on practitioners' perceived usefulness. The resulting ranking provides an empirical basis for selecting a concise set of core properties to be embedded into the guideline and its prompting strategy.

    \item \textbf{RQ3: To what extent does a guideline-based formulation tool affect the creation of explainability requirements compared to manual formulation?}\\
    Answering this research question allows us to evaluate the tool-supported formulation process as an intervention and to understand whether it changes outcomes relative to manual work.

    \begin{itemize}
        \item \textbf{RQ3.1: To what extent does the tool affect the \textit{efficiency} of creating explainability requirements compared to manual formulation?}\\
        This question focuses on resource usage---in particular, time effort---when requirements engineers formulate explainability requirements with and without tool support.

        \item \textbf{RQ3.2: To what extent does the tool affect the \textit{perceived quality} of created explainability requirements compared to manual formulation?}\\
        This question targets perceived requirement quality, operationalized through independent developer assessments of implementability and formulation quality of manually vs.\ tool-supported requirements.
    \end{itemize}
\end{itemize}

Figure~\ref{fig:overview-study} summarizes our sequential study design in four phases. Overall, the goal is to identify theoretically grounded, empirically prioritized, and practically applicable properties for formulating well-formulated explainability requirements, and to operationalize and evaluate them in a guideline-based formulation tool.

\textit{Phase 1: Elicitation of candidate properties.}
We collected potentially relevant quality properties through a structured literature review and complementary interviews with software developers.

\textit{Phase 2: Prioritization of properties.}
We conducted a quantitative online survey (\(n = 20\) complete responses) to assess the perceived usefulness of the elicited properties for formulating explainability requirements, including Likert ratings and Top-5/Bottom-5 selections.

\textit{Phase 3: Guideline derivation.}
We synthesized the prioritized properties into a concise formulation guideline that defines core quality properties and provides concrete formulation instructions (also used to steer LLM-based generation and revision).

\textit{Phase 4: Validation and tool evaluation.}
We evaluated the guideline and its tool-based operationalization through (i) a task-based study with requirements engineers to assess formulation efficiency and (ii) a subsequent independent online evaluation with software developers (\(n = 18\)) to assess perceived implementability and formulation quality of the resulting requirements, leading to minor refinements.

\subsection{Data Analysis}
\label{sec:data-analysis}

This section summarizes how we analyzed the data collected across the four study phases and the tool evaluation.

\subsubsection{Metrics}
\label{sec:metrics}

We used descriptive statistics and inferential tests aligned with the measurement scales of each outcome.

For efficiency (RQ3.1), we measured formulation time per requirement in seconds. Since times were right-skewed, we analyzed them with a linear mixed-effects model on log-transformed times to account for repeated measures (participants created multiple requirements and each requirement was produced in both conditions). As a sensitivity analysis, we additionally report a paired Wilcoxon signed-rank test on item-level pairs.

For perceived requirement quality (RQ3.2), participants rated implementability and formulation quality on an ordinal Likert scale. For each participant and condition, we aggregated ratings by taking the median across the relevant items and compared conditions with two-sided Wilcoxon signed-rank tests. We report the rank-biserial correlation ($r_{rb}$) as an effect size for paired ordinal comparisons and the Hodges--Lehmann estimator of the paired median difference ($\Delta_{HL}$) for $\mathrm{Tool} - \mathrm{Manual}$. For families of tests across requirement types, we applied Holm correction.

We also analyzed preference choices (manual vs.\ tool vs.\ tie) using descriptive counts at participant and item level. Optional open-ended feedback was coded against the guideline property categories (e.g., clarity, ambiguity, understandability) to connect qualitative feedback to the guideline.

We also analyzed ratings of how helpful the original app review was as additional context for implementation using descriptive distributions and medians (five-point ordinal scale).

\subsubsection{Types of Explainability Requirements}
\label{sec:types-explainability-reqs}

Our process distinguishes three types of explainability requirements, reflecting different ways to address an identified explanation need (captured from app reviews~\cite{obaidi2026multigolddataset}). The selected type shapes both requirement templates and the subsequent quality checks:

\begin{itemize}
    \item \textit{Direct User Reply.}
    The requirement specifies a direct response to the user, for example through a support or feedback channel.

    \item \textit{In-App Explanation.}
    The requirement specifies explanatory information provided to the user within the application, for example through tooltips, contextual help, or tutorials.

    \item \textit{System Adaptation.}
    The requirement specifies an adaptation of the system to prevent or reduce the explanation need, for example by changing functionality, workflows, or interface elements.
\end{itemize}

The tool supports selecting the type manually or determining it automatically via an LLM-based classification prompt. The resulting type is also used for reporting outcomes per type in addition to overall results.

\section{Guideline Derivation}
\label{sec:guideline-derivation}

This section describes how we derived the guideline by separating methodology from results. All studies involved participants aged 18 or older, surveys were anonymous, and participants were required to have relevant experience as specified below.

\subsection{Method}
\label{sec:guideline-method}

\subsubsection{Property Elicitation}
\label{sec:property-elicitation-method}

The goal of this phase (RQ1) was to elicit candidate quality properties for formulating well-formulated explainability requirements. We combined a structured literature review with developer interviews. 

\paragraph{Structured literature review}
We conducted a structured literature review to identify (i) general quality properties of requirements, with a focus on non-functional requirements, and (ii) properties discussed for explainability-related requirements. We searched Google Scholar, IEEE Xplore, and SpringerLink, and additionally used the AI-supported literature search tool Consensus. To broaden the search space beyond initial results, we applied backward and forward snowballing by systematically screening references of included papers and papers citing them.

We screened full texts against predefined inclusion and exclusion criteria (Table~\ref{tab:guideline-slr-criteria}). For each included source, we extracted reported requirement quality properties and consolidated them into a normalized candidate list by harmonizing terminology and merging synonyms where appropriate.

\begin{table}[htbp!]
\centering
\footnotesize
\caption{Inclusion and exclusion criteria for the structured literature review.}
\label{tab:guideline-slr-criteria}
\begin{tabularx}{\columnwidth}{@{}p{1.0cm}X@{}}
\toprule
\multicolumn{2}{@{}l}{\textbf{Inclusion criteria}} \\
\midrule
\textbf{I1} & The work addresses quality criteria of good requirements. \\
\textbf{I2} & The work addresses non-functional requirements or explainability. \\
\midrule
\multicolumn{2}{@{}l}{\textbf{Exclusion criteria}} \\
\midrule
\textbf{E1} & The publication is not accessible (open access or via university resources). \\
\textbf{E2} & The publication is not written in English or German. \\
\bottomrule
\end{tabularx}
\end{table}

\paragraph{Developer interviews}
To complement the literature-based candidate properties with practice perspectives, we conducted semi-structured interviews with three software developers. They had professional development experience. Interview topics covered typical weaknesses in real-world requirements, recurring formulation problems, and which aspects developers consider important when implementing requirements.

Interview statements were analyzed by mapping mentioned problems to requirement quality properties. When participants described weaknesses (e.g., vague wording), we translated these into positively phrased properties (e.g., unambiguity). We then compared the interview-derived properties to the literature-based list to identify confirmations and potential additions.

\subsubsection{Property Prioritization Survey}
\label{sec:property-prioritization-method}

This phase (RQ2) prioritized the elicited candidate properties to identify which ones software developers consider most useful for formulating explainability requirements. The survey was conducted online and restricted to participants with practical experience related to requirements.

\paragraph{Participants and recruitment}
We recruited participants for an online survey. Inclusion criteria were practical experience with requirements in software development, such as writing, implementing, testing, analyzing, or reviewing requirements. Participants provided demographic information and self-reported experience across common requirements-related activities to characterize the sample.

\paragraph{Survey instrument and procedure}
The survey started with demographic questions and questions about participants' experience with requirements. As input, we used the candidate property set derived from the literature review and interviews. Initially, 24 potential properties were available. Before running the survey, we consolidated the list to ensure properties were sufficiently distinct and could be rated independently. We removed three properties due to limited separability: changeability, level of detail, and simplicity. Simplicity overlapped conceptually with understandability and precision, and level of detail overlapped with completeness. Changeability mainly targets documentation and change processes rather than formulation quality.

The resulting 21 properties were rated on a 7-point Likert scale (1 = not useful at all, 7 = very useful) regarding their usefulness for formulating explainability requirements. Each property was presented with a short, fixed definition to reduce interpretation differences. To enforce relative priorities beyond generally high Likert agreement, participants additionally selected the five most important and five least important properties. The average completion time was about 20 minutes.

\subsubsection{Guideline Construction}
\label{sec:guideline-construction-method}

This phase combined the empirical prioritization results with theoretical grounding to derive a concise guideline that can be applied consistently during formulation and can also steer LLM-based support.

We selected core properties based on two criteria: high usefulness in the Likert ratings and frequent selection in the Top-5 ranking. We then cross-checked the resulting shortlist against ISO/IEC/IEEE~29148:2018~\cite{ISOIECIEEE.2018} and the interview findings to ensure theoretical alignment and practical relevance.

For each selected property, we formulated a short definition and a concrete formulation instruction intended for use during requirement writing and for integration into prompts that steer LLM-based generation and revision. We excluded properties that are hard to operationalize as formulation guidance without knowing the concrete stakeholder audience and usage context. This includes, for example, correctness and audience orientation, as both depend on information that cannot be inferred reliably from the requirement text alone.

\subsection{Results}
\label{sec:guideline-results}

\subsubsection{Elicited Candidate Properties}
\label{sec:elicited-properties-results}

\paragraph{Literature review results}
Our search and screening process covered 70 papers, from which we included eight core sources that directly address requirement quality properties in the context of non-functional requirements and/or explainability-related requirements. ISO/IEC/IEEE~29148:2018~\cite{ISOIECIEEE.2018} was a central reference, as it defines widely used requirement quality properties such as unambiguity, completeness, consistency, and verifiability. Additional sources contributed further properties commonly discussed in requirements engineering, including traceability, audience orientation, and understandability. The literature review resulted in a consolidated list of candidate properties for subsequent validation and prioritization.

\paragraph{Interview results}
The three interviewed developers (aged 25--29) had 2--8 years of professional experience in application, web, and system development. Two participants reported working with requirements several times per week, and one reported daily work with requirements. Across interviews, participants highlighted recurring shortcomings in real requirements, including unclear phrasing, missing precision, limited verifiability, and insufficient level of detail. These observations aligned well with the literature-based properties. The interviews did not yield entirely new properties not already covered in the literature, but they provided concrete examples and confirmed the practical relevance of the candidate set.

\paragraph{Synthesis of literature review and interviews}
Overall, literature and interviews provided a consistent picture of relevant quality properties for explainability requirements. Table~\ref{tab:guideline-candidate-properties} summarizes the consolidated candidate properties and indicates whether a property was explicitly confirmed by interview statements. This candidate set served as input to the subsequent prioritization survey (RQ2).

\begin{table}[htbp!]
\centering
\small
\setlength{\tabcolsep}{3pt}
\renewcommand{\arraystretch}{1.08}
\caption{Elicited candidate properties and their sources (literature and interviews).}
\label{tab:guideline-candidate-properties}
\begin{tabularx}{\columnwidth}{@{}p{2.85cm}>{\raggedright\arraybackslash}X>{\centering\arraybackslash}p{1.4cm}@{}}
\toprule
\textbf{Property} & \textbf{Source} & \textbf{Interview} \\
\midrule
Changeability & \cite{Rupp.2009} & \\
Appropriateness & \cite{ISOIECIEEE.2018,chazette2020explainability,Rupp.2009,kohl2019explainability} & \\
Terminological clarity & \cite{chazette2020explainability,kohl2019explainability} & \xmark \\
Level of detail & \cite{Pohl.2005} & \xmark \\
Unambiguity & \cite{ISOIECIEEE.2018,Rupp.2009,Montgomery.2022,Heck.2018} & \xmark \\
Uniformity & \cite{ISOIECIEEE.2018,Rupp.2009,kohl2019explainability} & \xmark \\
Clarity & \cite{ISOIECIEEE.2018,chazette2020explainability,Rupp.2009,kohl2019explainability} & \xmark \\
Consistency & \cite{Rupp.2009,deters2024qualitymodel,Heck.2018} & \\
Correctness & \cite{ISOIECIEEE.2018,chazette2020explainability,Rupp.2009,chazette2022explainable,deters2024qualitymodel,Montgomery.2022,Heck.2018} & \\
Readability & \cite{withall2021readability} & \\
Traceability & \cite{ISOIECIEEE.2018,Rupp.2009,chazette2022explainable,deters2024qualitymodel,Montgomery.2022,Heck.2018} & \\
Necessity & \cite{ISOIECIEEE.2018,Rupp.2009,kohl2019explainability} & \\
Objectivity & \cite{kohl2019explainability} & \xmark \\
Feasibility & \cite{ISOIECIEEE.2018,kohl2019explainability,Heck.2018} & \\
Compliance & \cite{ISOIECIEEE.2018,Rupp.2009,kohl2019explainability,Heck.2018} & \xmark \\
Simplicity & \cite{chazette2020explainability,kohl2019explainability} & \\
Singularity & \cite{ISOIECIEEE.2018} & \\
Structuredness & \cite{ISOIECIEEE.2018,Rupp.2009} & \\
Verifiability & \cite{ISOIECIEEE.2018,chazette2020explainability,Rupp.2009,kohl2019explainability,chazette2022explainable,Montgomery.2022,Heck.2018} & \xmark \\
Validatability & \cite{ISOIECIEEE.2018,kohl2019explainability,chazette2022explainable,Heck.2018} & \\
Understandability & \cite{ISOIECIEEE.2018,chazette2020explainability,kohl2019explainability,chazette2022explainable,deters2024qualitymodel,Montgomery.2022} & \xmark \\
Completeness & \cite{ISOIECIEEE.2018,chazette2020explainability,Rupp.2009,kohl2019explainability,Montgomery.2022,Heck.2018} & \\
Maintainability & \cite{Rupp.2009,chazette2022explainable} & \xmark \\
Audience orientation & \cite{Rupp.2009,kohl2019explainability,chazette2022explainable,deters2024qualitymodel,Heck.2018} & \xmark \\
\bottomrule
\end{tabularx}
\end{table}

\subsubsection{Prioritized Properties}
\label{sec:prioritized-properties-results}

A total of 34 participants started the survey and 20 completed it. All completers reported practical experience with requirements. Most participants reported implementing requirements (19/20) and creating or formulating requirements (17/20), while 9/20 reported testing requirements. Further activities included analyzing or checking requirements (6/20) and evaluating requirements (5/20). One participant reported adapting requirements in the context of traceability or change processes. The average age of completers was 27.4 years (range: 23--42).

Across the 21 properties, Likert ratings showed generally high agreement, which is expected because many of the properties are established requirement quality criteria. The highest ratings were observed for unambiguity, clarity, understandability, completeness, and correctness. Because these overall high ratings make fine-grained differentiation difficult, we relied on the forced ranking step to capture relative priorities.

The Top-5 selections reinforced the pattern and provided a clearer ordering. The most frequently selected properties among the Top-5 included correctness, unambiguity, completeness, and clarity (Figure~\ref{fig:ranking_pos_counts}). These Top-5 results served as the primary empirical input for selecting the core properties of the final guideline reported in the next section, while the sample size limits the strength of the prioritization evidence.

\begin{figure}[htbp!]
    \centering
    \includegraphics[width=0.95\columnwidth]{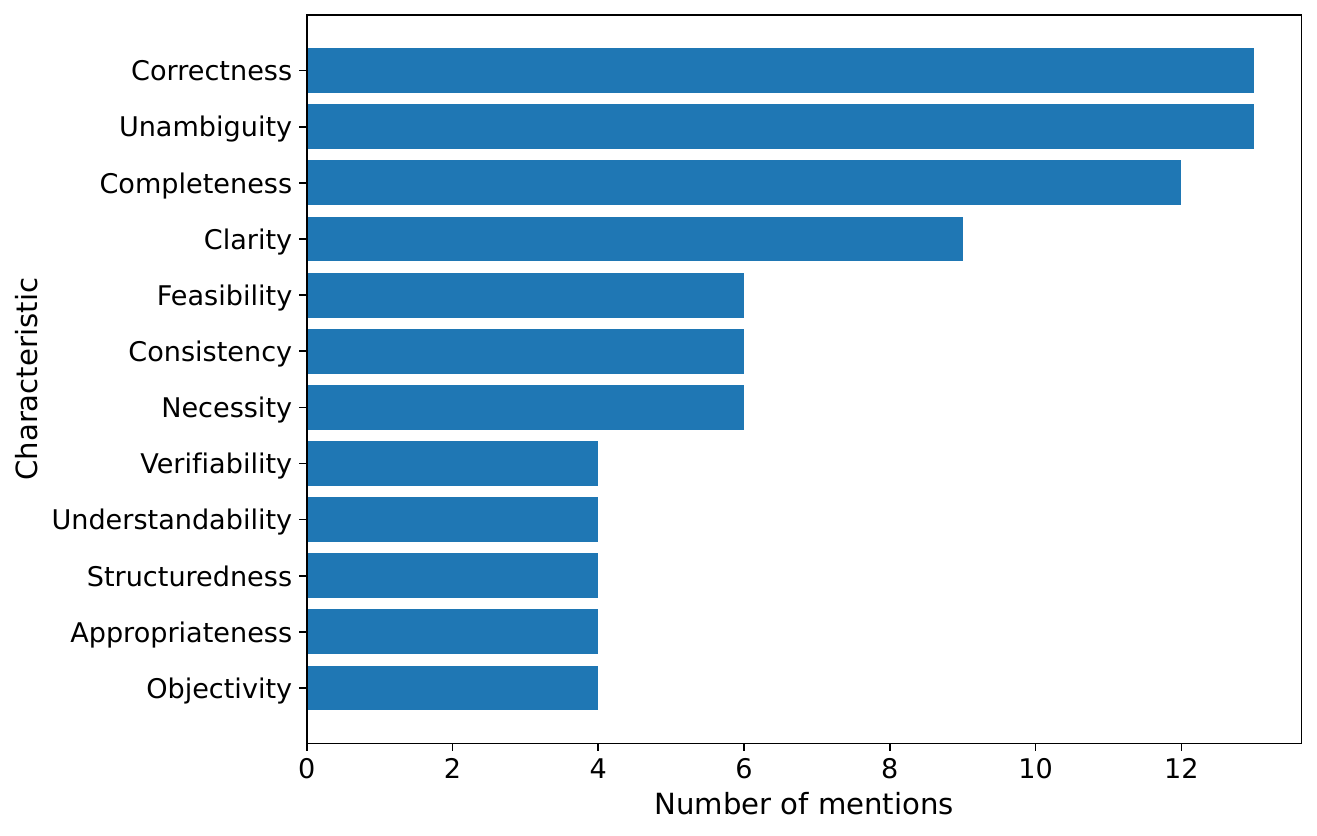}
    \caption{Most frequently selected Top properties (selected at least four times).}
    \label{fig:ranking_pos_counts}
\end{figure}

\subsubsection{Final Guideline}
\label{sec:final-guideline-results}

The final guideline consists of ten core properties selected as particularly relevant for formulating well-formulated explainability requirements: unambiguity, clarity, understandability, verifiability, readability, structuredness, singularity, objectivity, terminological clarity, and completeness. For each property, the guideline provides a short definition and a concrete formulation instruction that can be applied during requirement writing and can be embedded into LLM prompts to steer generation and iterative revision.

Table~\ref{tab:krahl-guideline} presents the full guideline. We intentionally did not include correctness among the core properties, although it was highly prioritized, because assessing correctness requires access to the actual stakeholder need and usage context rather than only the textual formulation.

\begin{table*}[htbp!]
    \centering
    \scriptsize
    \renewcommand{\arraystretch}{1.2}
    \setlength{\tabcolsep}{4pt}
    \caption{Guideline for formulating well-formulated explainability requirements: definitions and concrete formulation instructions intended to be applied during requirement writing and in LLM-based support.}
    \label{tab:krahl-guideline}
    \begin{tabularx}{\textwidth}{@{}p{2.0cm}p{8.0cm}X@{}}
        \toprule
        \textbf{Property} & \textbf{Definition} & \textbf{Formulation instruction for an LLM} \\
        \midrule
        \textbf{Unambiguity} &
        A requirement is unambiguous if it allows only one interpretation and contains no ambiguous wording or terms. &
        Use precise, measurable terms and avoid vague expressions such as ``fast'', ``appropriate'', or ``intuitive''. \\
        \midrule
        \textbf{Clarity} &
        A requirement is clear if its intent is expressed without room for interpretation and the intended meaning is immediately apparent. &
        Write short, active sentences and avoid nested clauses or unnecessarily complex structures. \\
        \midrule
        \textbf{Understandability} &
        A requirement is understandable if it is readable for all relevant stakeholders and comprehensible in the intended context. &
        Use plain language, avoid unnecessary jargon, and keep terminology consistent. \\
        \midrule
        \textbf{Verifiability} &
        A requirement is verifiable if objective criteria or tests exist to determine whether it is satisfied, including for qualitative NFRs such as explainability. &
        Specify measurable criteria (e.g., time, counts, percentages) and avoid non-measurable terms such as ``pleasant'' or ``user-friendly''. \\
        \midrule
        \textbf{Readability} &
        A requirement is readable if it is fluent and easy to process linguistically. &
        Use short sentences and simple wording. Avoid unnecessary complexity. \\
        \midrule
        \textbf{Structuredness} &
        A requirement is structured if it follows a clear documentation structure, for example a template, sectioning, or hierarchy. &
        Follow the selected template and avoid unstructured prose. \\
        \midrule
        \textbf{Singularity} &
        A requirement is singular if it describes exactly one aspect or condition and cannot be split into multiple sub-requirements. &
        Split multiple goals into separate requirements. Avoid enumerations or long ``and'' chains. \\
        \midrule
        \textbf{Objectivity} &
        A requirement is objective if it is based on checkable facts and contains no subjective judgments. &
        Avoid evaluative adjectives such as ``good'' or ``attractive''. Support claims with verifiable facts. \\
        \midrule
        \textbf{Terminological clarity} &
        A requirement has terminological clarity if used terms are defined or referenced to prevent misunderstandings. &
        Use clearly defined terms. Expand abbreviations at first use. Avoid placeholders such as ``suitable'' or ``relevant''. \\
        \midrule
        \textbf{Completeness} &
        A requirement is complete if it contains all necessary information for implementation and verification, including conditions and explanation-related criteria. &
        Explicitly state actor, trigger, system behavior, and expected outcome to avoid gaps. \\
        \bottomrule
    \end{tabularx}
\end{table*}

\section{Tool Development and Evaluation}
\label{sec:tool-evaluation}

This section evaluates a guideline-driven formulation tool as an intervention in the creation of explainability requirements. All studies involved participants aged 18+. Surveys were anonymous. Participants were required to have relevant professional experience, as specified in the respective study descriptions.

\subsection{Tool Overview}
\label{sec:tool-overview}

The tool operationalizes the guideline from Section~\ref{sec:guideline-derivation} to support the formulation of explainability requirements. The goal is not fully automated requirement generation. The goal is to keep the user in control of content and level of detail while supporting more consistent and guideline-aligned formulation.

The tool embeds the guideline directly into a three-step workflow:
\begin{enumerate}
    \item \textbf{Generation.} The user provides an explanation need as input. This can be a full app review or an extracted need statement. The tool generates an initial explainability requirement based on this context and the selected guideline properties.
    \item \textbf{Quality checks.} The generated requirement is assessed along all guideline properties. The tool highlights which properties are met and where concrete improvements are needed.
    \item \textbf{Revision.} The user iteratively revises the requirement. The tool supports this through targeted quick fixes and rephrasing, as well as manual edits. A custom prompt option allows focused modifications without rewriting the requirement from scratch.
\end{enumerate}

The tool was designed as a formulation aid rather than an autonomous requirements generator. This design choice reflects that correctness and stakeholder alignment cannot be inferred reliably from the input text alone. The LLM therefore supports drafting, checking, and revision, while the requirements engineer remains responsible for interpreting the explanation need and validating the resulting requirement.

To reflect different ways of addressing an explanation need, the tool supports three requirement types. The type can be selected manually or determined automatically by an LLM-based classification step. The three types are \textit{Direct User Reply}, \textit{In-App Explanation}, and \textit{System Adaptation} (Section~\ref{sec:types-explainability-reqs}). 

The input is expected to represent an identified explanation need. In a broader explainability management process, app reviews or other user feedback (e.g., \cite{anders2023userfeedback}) can first be filtered and classified, for example with taxonomies of explanation needs. If an input is vague or admits multiple interpretations, it should be inspected by a requirements engineer before formulation. The tool can support wording and structure, but it cannot reconstruct missing context or determine the original user's intent with certainty.

\paragraph{LLM integration and prompt design}
The tool uses GPT-4o-mini via the OpenAI API. At the time of development, GPT-4o-mini provided a strong balance between output quality and low latency with stable API access, which is important for interactive, real-time tooling. All API calls are encapsulated in a service layer to support model changes with low implementation effort. The prompt design follows a modular structure:
\begin{itemize}
    \item A system instruction that injects the selected guideline properties as compact formulation rules.
    \item A per-property quality-check prompt that returns a score from 1 to 4 and short improvement suggestions.
    \item A classification prompt that assigns the explanation need to one of the three requirement types.
    \item Quick-fix prompts that aggregate unmet properties into a short correction instruction to revise the requirement while preserving its intended meaning.
\end{itemize}
We treat the prompt design as one operationalization of the guideline rather than as an optimized prompting strategy. We keep the paper focused on the study outcomes. All prompts, templates, and parameter settings are available at \href{https://figshare.com/s/e8b2063a271659c35db7}{figshare}.

\paragraph{Tool functions and outputs}
The tool provides configuration of guideline properties and templates, generation and revision functions, and a quality-check dashboard. Each guideline property is scored from 1 to 4. The tool also computes an overall quality score by normalizing the sum of property scores to a 0 to 100 range. The interface follows the process structure with three main areas for guideline configuration, requirement formulation, and quality assessment. Figure~\ref{fig:tool-screenshot} shows an example screen of the tool.

\begin{figure}[htbp]
\centering
\includegraphics[width=1.00\columnwidth]{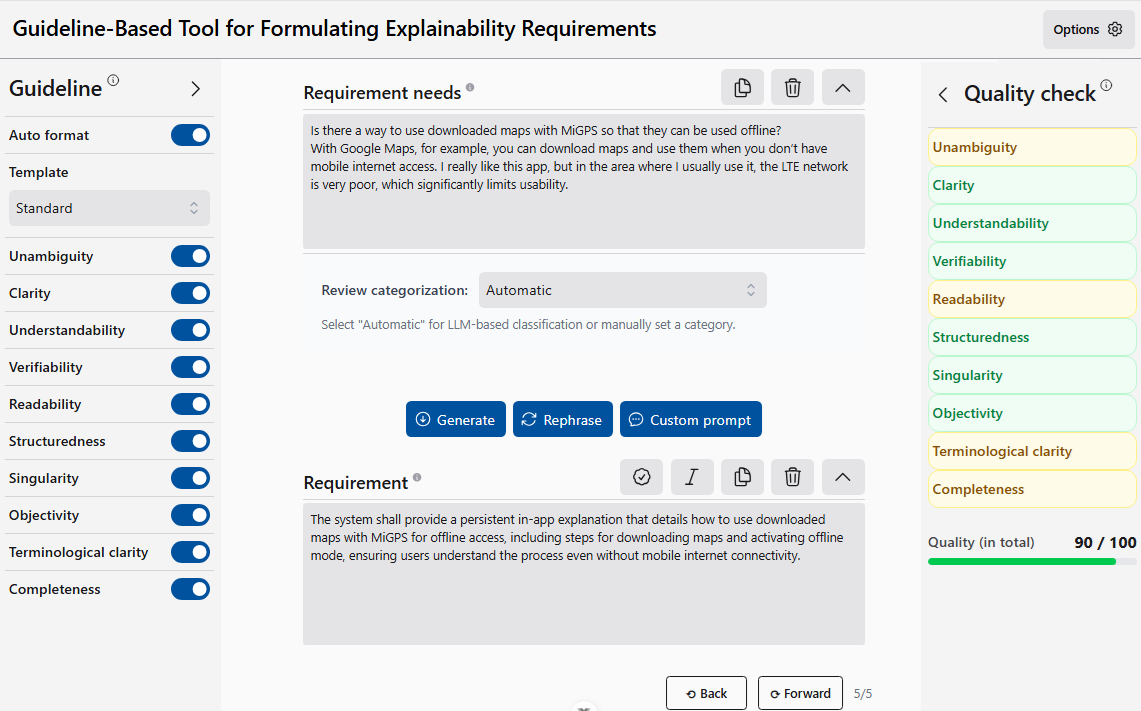}
\caption{Screenshot of the guideline-driven formulation tool for explainability requirements.}
\label{fig:tool-screenshot}
\end{figure}

Since LLM outputs are non-deterministic and guideline properties can interact, the tool does not guarantee that all properties are fully satisfied in every iteration. Instead, it supports systematic improvement by making strengths and weaknesses explicit and by enabling targeted revisions.

\subsection{Evaluation Method}
\label{sec:tool-evaluation-method}

\subsubsection{Task-Based Study}
\label{sec:task-based-study-method}

\paragraph{Participants and recruitment}
We recruited six practitioners who work with requirements in practice. All participants were male and between 26 and 31 years old (mean 27.5). Their professional experience ranged from 2 to 8 years (mean 4.9). Three participants worked as requirements engineers, two as software engineers, and one in a software/system house context. Five participants reported frequent or very frequent work with requirements, while one reported less frequent involvement. We did not require domain-specific expertise, because the study focused on formulation support rather than domain-specific correctness.

\paragraph{Procedure and measures}
The moderated online task-based study used a within-subject design and included a short tool introduction and practice runs before the timed tasks. Each participant formulated three explainability requirements manually with the guideline and then formulated three additional requirements with tool support. The six requirements per participant covered the three requirement types described in Section~\ref{sec:types-explainability-reqs}. The tasks were based on app-review-derived explanation needs~\cite{obaidi2026multigolddataset} and provided lightweight context for formulation. We recorded the formulation time in seconds for each requirement. We also collected short qualitative notes about the workflow and perceived comprehensibility during the session.

\subsubsection{Online Study}
\label{sec:online-quality-method}

\paragraph{Participants and recruitment}
Of 35 participants who started the survey, 18 completed it. Participants were 24--35 years old (mean 28 years) and reported early to mid-level professional experience in software development.

\paragraph{Instrument and procedure}
Participants evaluated the requirements produced in the task-based study. For each item, we presented the manually formulated and the tool-supported version. Participants rated each version on a 7-point Likert scale from 1 (strongly disagree) to 7 (strongly agree), recoded to $-3$ to $+3$ for analysis:
\begin{itemize}
    \item This formulation would help me implement the requirement (implementability).
    \item I consider this requirement well formulated (formulation quality).
\end{itemize}
Participants could optionally provide open feedback. We also captured preference choices between manual and tool-supported requirements (manual, tool, tie). In addition, we asked how helpful the original app review was as additional context for implementation using a 5-point Likert scale from $-2$ to $+2$.

\subsection{Results}
\label{sec:tool-evaluation-results}

\subsubsection{Efficiency Results}
\label{sec:efficiency-results}

\begin{table}[htbp!]
\centering
\footnotesize
\caption{Average formulation time per requirement type and savings with tool support.}
\label{tab:efficiency-time}
\begin{tabularx}{\columnwidth}{l *{4}{>{\centering\arraybackslash}X}}
\toprule
\textbf{Requirement type} &
\textbf{Manual (mm:ss)} &
\textbf{Tool (mm:ss)} &
\textbf{Savings (\%)} &
\textbf{Savings (s)} \\
\midrule
All requirements &
01:51 &
01:25 &
23.5\% &
26 \\
\midrule
\textit{Direct User Reply} &
01:46 &
01:15 &
29.3\% &
31 \\
\textit{In-App Explanation} &
01:17 &
01:03 &
18.5\% &
14 \\
\textit{System Adaptation} &
02:31 &
01:58 &
21.9\% &
33 \\
\bottomrule
\end{tabularx}
\end{table}

Across all requirements, the average formulation time decreased from 1:51~min:s (manual) to 1:25~min:s (tool), which corresponds to a reduction of 23.5\% and an average saving of 26~s per requirement (Table~\ref{tab:efficiency-time}). A mixed-effects model on log-transformed times showed a significant tool effect ($\beta=-0.421$, two-sided $p=0.049$). This corresponds to a Tool/Manual multiplier of 0.66 with a 95\% CI of 0.43 to 1.00 and an estimated reduction of about 34\%. A paired Wilcoxon signed-rank test on item-level pairs ($n=18$) also indicated a significant difference ($p=0.021$) with a strong effect size ($r_{rb}=0.620$).

Descriptively, reduced formulation times appeared across all requirement types. The largest reduction was observed for \textit{Direct User Reply}. Due to the small number of item pairs per type ($n=6$), we do not report separate significance tests per type.

\subsubsection{Perceived Quality Results}
\label{sec:quality-results}

\begin{table*}[htbp!]
\centering
\scriptsize
\caption{Median (MD) and interquartile range in brackets for participant-level aggregated ratings ($-3$ to $+3$) on implementability and formulation quality.}
\label{tab:quality-medians}
\begin{tabularx}{\textwidth}{
l
>{\raggedleft\arraybackslash}X
>{\raggedleft\arraybackslash}X
>{\raggedleft\arraybackslash}X
>{\raggedleft\arraybackslash}X
>{\raggedleft\arraybackslash}X
>{\raggedleft\arraybackslash}X
>{\raggedleft\arraybackslash}X
>{\raggedleft\arraybackslash}X
}
\toprule
\textbf{Median} & \multicolumn{2}{c}{\textbf{Direct User Reply}}
& \multicolumn{2}{c}{\textbf{In-App Explanation}}
& \multicolumn{2}{c}{\textbf{System Adaptation}}
& \multicolumn{2}{c}{\textbf{Overall}} \\
\cmidrule(lr){2-3} \cmidrule(lr){4-5} \cmidrule(lr){6-7} \cmidrule(lr){8-9}
& \textbf{Manual} & \textbf{Tool}
& \textbf{Manual} & \textbf{Tool}
& \textbf{Manual} & \textbf{Tool}
& \textbf{Manual} & \textbf{Tool} \\
\midrule
Implementability
& 2 [1.38] & 1 [1]
& 1 [1] & 2 [1.5]
& 1.5 [1.38] & 2 [1.38]
& 1.5 [1] & 2 [1] \\
Formulation quality
& 1.25 [1.88] & 1 [0.5]
& 1 [1.88] & 1.5 [1]
& 1 [0.88] & 1 [0.88]
& 1 [1] & 1 [0.88] \\
\bottomrule
\end{tabularx}
\end{table*}

Table~\ref{tab:quality-medians} summarizes implementability and formulation quality ratings. On the overall level, we found no statistically significant differences between manual and tool-supported requirements. For implementability, the medians were $MD_{\mathrm{manual}}=1.5$ and $MD_{\mathrm{tool}}=2$ with no significant difference (Wilcoxon $p=0.351$). For formulation quality, both conditions had the same median ($MD=1$) with no significant difference ($p=0.802$). The Hodges--Lehmann estimator was $\Delta_{HL}=0$ for both outcomes.

When analyzing requirement types separately, none of the comparisons remained significant after Holm correction (all $p_{\mathrm{Holm}} \ge 0.247$). The strongest trend appeared for \textit{In-App Explanation} in implementability ($p=0.082$, $r_{rb}=0.50$, $\Delta_{HL}=0.5$).

Preference choices showed a descriptive slight tendency toward tool-supported requirements. For implementability, participants preferred the manual version 107 times and the tool-supported version 124 times, with 93 ties. On the participant level, six participants preferred manual more often, eight preferred tool more often, and four had an overall tie. On the item level, five items favored manual, eight favored tool, and five were ties. For formulation quality, the preference pattern was similar.

We received ten optional open comments. All comments could be mapped to the property categories used in the guideline. The most frequent themes were understandability (7 mentions) and clarity (5). Further mentions included unambiguity (3), simplicity (3), and readability (2). Single mentions addressed level of detail, singularity, structuredness, and completeness.

\subsubsection{Remaining Issues in Tool-Supported Requirements}
\label{sec:remaining-formulation-issues}

To better understand the limits of tool-supported formulation, we inspected the two tool-supported requirements with the lowest median ratings. We report translated excerpts from the generated requirements rather than the full texts to save space.

\textbf{ER\,11} had the lowest median rating for implementability (median 1.0). The tool-supported requirement stated that the system should ``provide in-app explanations to make the user interface of the mathematics app easier to understand'' and mentioned possible solutions such as ``tooltips, onboarding, clear error messages, and permission rationales''. However, the requirement did not specify which parts of the interface were difficult to understand, when explanations should be shown, or how satisfaction could be verified. The main remaining issues concerned completeness, singularity, and verifiability.

\textbf{ER\,17} had the lowest median rating for formulation quality (median 0.5). The tool-supported requirement asked for ``error-free quantity information for juices in the shopping list'' and added that future development could follow the iOS version because it was perceived as more advanced. This mixes a functional correctness issue with an unclear platform comparison and does not clearly express an explainability-related system behavior. The main remaining issues concerned clarity, objectivity, and verifiability.

These examples show that tool support can produce structured formulations while still leaving important formulation issues unresolved. Requirements engineers therefore still need to interpret the explanation need, check whether relevant context is missing, and decide whether the selected requirement type and acceptance criteria are appropriate.
\subsubsection{Usefulness of App Reviews}
\label{sec:review-usefulness-results}

Figure~\ref{fig:review-usefulness} shows how developers rated the helpfulness of the original app review as additional context for implementation. We collected this rating only for three representative items, one per requirement type in the category-grouped analysis order: item~6 (\textit{Direct User Reply}), item~12 (\textit{In-App Explanation}), and item~18 (\textit{System Adaptation}). Ratings were distributed across all response options. Neutral and slightly positive ratings occurred more often than strongly negative or strongly positive ones. This supports treating app reviews as useful but incomplete context: they can inform formulation in some cases, but they should not replace a self-contained requirement or human interpretation of the explanation need.

\begin{figure}[htbp]
    \centering
    \includegraphics[width=0.95\linewidth]{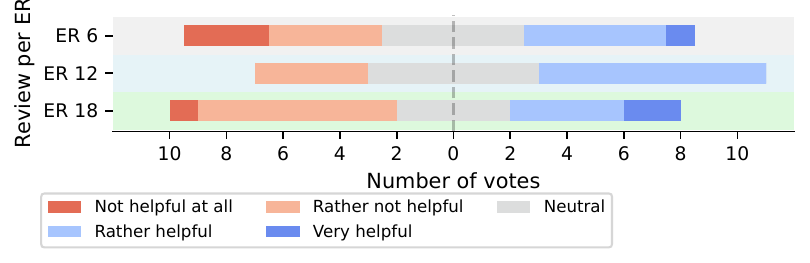}
    \caption{Developers' ratings of the helpfulness of app reviews as additional context for implementing explainability requirements ($n=18$).}
    \label{fig:review-usefulness}
\end{figure}

\section{Discussion}
\label{sec:discussion}

In the following, we answer the research questions, present threats to validity, and interpret the results.

\subsection{Answers to the Research Questions}
\label{sec:answers-rqs}

\textbf{RQ1: Which properties characterize well-formulated explainability requirements from research and practice?}\\
The literature review and developer interviews yielded a consolidated set of candidate requirement quality properties, including established criteria such as unambiguity, completeness, and verifiability, complemented by practice-relevant properties such as readability, structuredness, and terminological clarity. Interviews supported the relevance of these properties by pointing to recurring issues in real requirements, such as vague wording, missing test criteria, and insufficient detail.

\textbf{RQ2: Which of these properties are perceived by software developers as most useful for formulating explainability requirements?}\\
Developers rated most properties highly, while the Top-5 ranking provided clearer differentiation. The most frequently selected Top properties were correctness, unambiguity, completeness, and clarity, which informed the core selection used to derive the guideline. 

\textbf{RQ3: To what extent does a guideline-based formulation tool affect the creation of explainability requirements compared to manual formulation?}\\
Formulation time was 23.5\% lower with tool support (26 seconds per requirement) and the time difference was significant in both the mixed-effects model and the paired Wilcoxon test. Perceived implementability and formulation quality did not differ significantly between manual and tool-supported requirements, while preferences showed a slight tendency toward tool-supported variants.

\subsection{Interpretation}
\label{sec:interpretation}

Our findings complement prior work on explainability requirements by addressing a later step in the process. Existing taxonomies and elicitation approaches help identify and structure explanation needs, while our workflow focuses on turning such needs into formulated requirements. In this sense, the contribution is not another taxonomy or elicitation method, but a lightweight operationalization of requirement quality properties for the formulation step.

The results indicate that the main value of using an LLM in this context comes from methodical steering rather than from generation alone. The guideline turns widely discussed requirement quality principles into a small, operational set of checks and writing rules that can be applied consistently. The tool then embeds these rules into an iterative workflow that supports drafting, diagnosing weaknesses, and revising toward clearer, more testable requirements.

At the same time, our results qualify expectations about LLM-based support in requirements engineering. The study does not show a significant improvement in perceived implementability or formulation quality. Instead, the observed benefit is primarily efficiency-related: formulation time was lower with tool support while perceived quality did not differ significantly from manual formulation. This may enable faster iteration cycles, but it should not be interpreted as evidence that the current tool reliably produces higher-quality requirements than manual formulation. Human validation remains central, especially for ensuring that a requirement matches the intended stakeholder need and usage context.

The three requirement types provide a useful framing for practice. They make explicit that an explanation need can be addressed by a direct user reply, an explanation within the application, or an adaptation of the system that reduces the explanation need. This helps requirements engineers avoid prematurely committing to one solution form and supports clearer alignment between requirement intent and implementation strategy.

The remaining issues observed in the low-rated tool-supported requirements further show that structured wording is not sufficient. Some requirements still lacked specific explanation content, acceptance criteria, or a clearly delimited interpretation of the input need. ER\,17 also illustrates that tool support may preserve or rephrase inputs that are closer to functional correctness or platform comparison than to explainability. Together with the heterogeneous ratings of app reviews as implementation context, this indicates that user feedback can inform formulation, but should not replace a self-contained requirement or human interpretation of the explanation need. For practice, requirements engineers should therefore treat tool output as a review candidate rather than as a finished requirement, especially when the input review is vague, mixes concerns, or does not clearly express an explanation need.

For stakeholders, the most immediate benefit is more structured formulation support and potential time savings. Requirements engineers can produce and iterate on explainability requirements faster, and developers receive specifications whose ratings did not differ significantly from manually written ones. For organizations, this may support scaling explainability work across products and teams by standardizing how requirements are phrased and reviewed, while keeping accountability with humans rather than the model.

\subsection{Threats to Validity}
\label{sec:validiteat}

This subsection applies the threats to validity categories described by Wohlin et al.~\cite{wohlin2012experimentation}. We discuss construct, internal, conclusion, and external validity for the guideline derivation and tool-based evaluation.

\subsubsection{Construct Validity}
Our evaluation relies on subjective judgments for several constructs, including perceived implementability and formulation quality, which were captured via Likert scales. Participants may interpret these constructs differently, and the generally high ratings observed for many properties suggest a positivity tendency that can reduce contrast between conditions and between properties. In addition, the online evaluation asked participants to judge requirements outside a full project context. This limits how well ratings reflect real work situations in which implementation constraints and domain knowledge are available.

The tool also includes model-based quality checks and an aggregated score. These checks are not objective measures of requirement quality and may be unreliable for properties such as clarity and understandability. We therefore treat them as support for revision, not as primary outcome measures. Similarly, the study measures perceived quality rather than objective downstream quality, such as reduced implementation defects or test effort. Triangulating multiple data sources (literature, interviews, survey prioritization, task-based study time measures, and independent ratings) mitigates mono-method bias, but does not remove it.

\subsubsection{Internal Validity}
The task-based study used a within-subject design, which is sensitive to learning and order effects. Participants formulated requirements manually first and then with tool support, so part of the time reduction may be due to familiarization with the task. We captured qualitative notes during the task-based study to detect unusual behavior, but we cannot fully separate tool effects from practice effects.

Participant differences can also confound results. Variation in requirements experience, writing style, motivation, and familiarity with language models can influence both formulation time and how tool suggestions are applied. The small sample sizes in the task-based study ($n=6$) and online study ($n=18$) increase the influence of individual differences. The prioritization survey was also based on a small completed sample ($n=20$), which limits the strength of the empirical basis for selecting the guideline properties.

\subsubsection{Conclusion Validity}
Sample sizes limit statistical power, especially for analyses by requirement type. We therefore interpret type-specific patterns as trends rather than strong evidence. We used analysis methods aligned with measurement scales, including mixed-effects modeling for repeated time measurements and paired nonparametric tests for ordinal outcomes. We also reported effect sizes and applied Holm correction for families of type-wise comparisons. Despite these precautions, the results should be interpreted as indicative, not as definitive causal proof for all settings.

\subsubsection{External Validity}
Generalizability is limited by the participant pools and tasks. Participants were recruited via convenience sampling and may not represent the full diversity of industrial roles and organizational contexts. The evaluation covered 18 paired requirements and three requirement types, which constrains the range of domains, systems, and specification styles.

The results also depend on the specific guideline operationalization and the chosen language model (GPT-4o-mini) and prompt design. Changes in model versions, API behavior, or prompt wording may lead to different outcomes. While the overall workflow of guideline-driven generation, checking, and revision is model-agnostic, the measured effects should be revalidated when transferring the approach to other domains, organizations, languages, or model configurations. In addition, the approach depends on the quality of the input explanation need. If an app review or extracted need is vague, incomplete, or ambiguous, the tool may produce a well-structured requirement that is still incomplete or misaligned with the intended stakeholder need.

\subsection{Future Work}
\label{sec:future-work}

Future work can extend both the guideline and the tool-based workflow.

First, the guideline should be validated with larger and more diverse samples and across additional domains. This includes checking whether context-specific properties are needed and whether some properties can be measured more objectively. A related step is to study how reliably LLM-based quality checks reflect human judgments for each property and which properties are most difficult to assess automatically.

Second, the tool can be improved to better support professional specification work. Prompts and templates should be maintained as models evolve, and the workflow could benefit from more adaptive feedback that targets the weakest properties of a requirement. Another direction is to integrate domain knowledge and organizational sources, such as product documentation, decision logs, and design rationales, to reduce the risk of producing requirements that are well phrased but misaligned with the real stakeholder need.

Third, the approach should be evaluated in real industrial projects and over longer periods. Such studies can assess how the workflow affects communication across roles, rework rates, and downstream outcomes such as implementability, testability, and change effort. They can also examine how the tool fits into existing RE practices and governance, including traceability and review processes.

Finally, future studies should analyze outcomes at the level of individual guideline properties. Instead of only collecting global quality ratings, researchers can measure how strongly tool-supported requirements satisfy each property and how these property-level measures relate to perceived usefulness and acceptance. This would help identify which properties drive improvements and where additional support mechanisms are needed.

\section{Conclusion}
\label{sec:conclusion}

This paper investigated how to support the formulation of explainability requirements through a guideline-driven workflow and tool support. We followed a sequential approach that elicited candidate quality properties from a structured literature review and developer interviews, prioritized them in an anonymous online survey, and derived a concise guideline with actionable formulation instructions. We then operationalized the guideline in an LLM-supported tool and evaluated it in two studies: a task-based study with requirements engineers and an independent online study with software developers.

Our results suggest that the workflow can reduce formulation effort, while we found no evidence of lower perceived requirement quality in our study. In the task-based study, average formulation time was 23.5\% lower with tool support and the difference was significant. In the online study, manually written and tool-supported requirements did not differ significantly in implementability or formulation quality, with a descriptive slight preference tendency toward tool-supported variants. Developers rated app reviews as additional implementation context heterogeneously, which suggests that reviews can help in some cases but should not replace self-contained requirements.

Overall, the main benefit of LLM support in this setting comes from steering formulation with prioritized quality properties rather than unguided generation. The results should therefore be interpreted as evidence for efficiency-oriented support, not as proof of improved requirement quality. Future work should validate the guideline and workflow in larger industrial settings and examine how organizational knowledge sources affect correctness and downstream outcomes.

\section*{Acknowledgment}
This work was funded by the Deutsche Forschungsgemeinschaft (DFG, German Research Foundation) under Grant No. 470146331, project softXplainer (2026-2028).

\section*{Data Availability Statement}
\label{sec:datastatement}
All artifacts produced by or used in this study, including data, tool code, prompts, study materials, and analysis scripts, are publicly available on Zenodo~\cite{obaidi2026guidelineReqDataset}.

\bibliographystyle{IEEEtran}
\bibliography{references.bib}

@book{wohlin2012experimentation,
  year={2012},
  title={Experimentation in software engineering},
  publisher={Springer},
  author={Claes Wohlin and Runeson, Per and Höst, Martin and Ohlsson, Magnus C. and Regnell, Björn and Wesslén, Anders},
}

@article{chazette2020explainability,
  title={Explainability as a non-functional requirement: challenges and recommendations},
  author={Chazette, Larissa and Schneider, Kurt},
  journal={REJ},
  volume={25},
  number={4},
  year={2020},
  publisher={Springer}
}

@article{chazette2022explainable,
  title={Explainable software systems: from requirements analysis to system evaluation},
  author={Chazette, Larissa and Brunotte, Wasja and Speith, Timo},
  journal={REJ},
  volume={27},
  number={4},
  year={2022},
  publisher={Springer}
}

@inproceedings{unterbusch2023explanation,
  title={Explanation Needs in App Reviews: Taxonomy and Automated Detection},
  author={Unterbusch, Max and Sadeghi, Mersedeh and Fischbach, Jannik and Obaidi, Martin and Vogelsang, Andreas},
  booktitle={REW},
  year={2023},
  organization={IEEE}
}

@inproceedings{kohl2019explainability,
  title={Explainability as a non-functional requirement},
  author={K{\"o}hl, Maximilian A and Baum, Kevin and Langer, Markus and Oster, Daniel and Speith, Timo and Bohlender, Dimitri},
  booktitle={2019 IEEE 27th International Requirements Engineering Conference (RE)},
  pages={363--368},
  year={2019},
  organization={IEEE}
}

@inproceedings{droste2023designing,
  title={Designing end-user personas for explainability requirements using mixed methods research},
  author={Droste, Jakob and Deters, Hannah and Puglisi, Joshua and Kl{\"u}nder, Jil},
  booktitle={REW},
  year={2023},
  organization={IEEE}
}

@inproceedings{droste2024explanations,
  author={Droste, Jakob and Deters, Hannah and Obaidi, Martin and Schneider, Kurt},
  booktitle={2024 IEEE 32nd International Requirements Engineering Conference (RE)}, 
  title={Explanations in Everyday Software Systems: Towards a Taxonomy for Explainability Needs}, 
  year={2024},
  volume={},
  number={},
  pages={55-66},
  doi={10.1109/RE59067.2024.00016}
}

@inproceedings{deters2024pulse,
  title={On the Pulse of Requirements Elicitation: Physiological Triggers and Explainability Needs},
  author={Deters, Hannah and Droste, Jakob and Schneider, Kurt},
  booktitle={REFSQ Workshops},
  year={2024},
  publisher={CEUR Workshop Proceedings}
}

@InProceedings{deters2024qualitymodel,
author="Deters, Hannah and Droste, Jakob and Obaidi, Martin and Schneider, Kurt",
title="How Explainable Is Your System? Towards a Quality Model for Explainability",
booktitle="Requirements Engineering: Foundation for Software Quality",
year="2024",
publisher="Springer Nature Switzerland",
address="Cham",
pages="3--19",
isbn="978-3-031-57327-9"
}

@inproceedings{chazette2021exploring,
  title={Exploring explainability: a definition, a model, and a knowledge catalogue},
  author={Chazette, Larissa and Brunotte, Wasja and Speith, Timo},
  booktitle={RE},
  year={2021},
  organization={IEEE}
}

@INPROCEEDINGS{stapel2009flow,
  author={Stapel, Kai and Knauss, Eric and Schneider, Kurt},
  booktitle={2009 Collaboration and Intercultural Issues on Requirements: Communication, Understanding and Softskills}, 
  title={Using FLOW to Improve Communication of Requirements in Globally Distributed Software Projects}, 
  year={2009},
  volume={},
  number={},
  pages={5-14},
}

@inproceedings{ramos2021modeling,
  title={Modeling and evaluating personas with software explainability requirements},
  author={Ramos, Henrique and Fonseca, Mateus and Ponciano, Lesandro},
  booktitle={Human-Computer Interaction: 7th Iberoamerican Workshop, HCI-COLLAB 2021, Sao Paulo, Brazil, September 8--10, 2021, Proceedings 7},
  pages={136--149},
  year={2021}
}

@INPROCEEDINGS{chazette2022framework,
  author={Chazette, Larissa and Klös, Verena and Herzog, Florian and Schneider, Kurt},
  booktitle={RE},
  title={Requirements on Explanations: A Quality Framework for Explainability},
  year={2022},
  volume={},
  number={},
}

@inproceedings{obaidi2025automatingexplanationneedmanagement,
  author = {Obaidi, Martin and Voß, Nicolas and Droste, Jakob and Deters, Hannah and Herrmann, Marc and Fischbach, Jannik and Schneider, Kurt},
  title = {Automating Explanation Need Management in App Reviews: A Case Study from the Navigation App Industry},
  year = {2025},
  location = {Ottawa, Canada},
  booktitle = {ICSE-SEIP'25}
}

@InProceedings{obaidi2025mood,
author="Obaidi, Martin and Droste, Jakob and  Deters, Hannah and Herrmann, Marc and Kl{\"u}nder, Jil and Schneider, Kurt",
title="Do Users' Explainability Needs in Software Change with Mood?",
booktitle="REFSQ'25",
year="2025",
}

@InProceedings{obaidi2025appKnowledge,
author="Obaidi, Martin and Fischbach, Jannik and Herrmann, Marc and Deters, Hannah and Droste, Jakob and Kl{\"u}nder, Jil and Schneider, Kurt",
title="How Does Users' App Knowledge Influence the Preferred Level of Detail and Format of Software Explanations?",
booktitle="REFSQ'25",
year="2025",
}

@inproceedings{brunotte2023context,
  title={Context, Content, Consent-How to Design User-Centered Privacy Explanations (S).},
  author={Brunotte, Wasja and Droste, Jakob and Schneider, Kurt},
  booktitle={SEKE},
  pages={86--89},
  year={2023}
}

@article{adadi2018peeking,
  title={Peeking inside the black-box: a survey on explainable artificial intelligence (XAI)},
  author={Adadi, Amina and Berrada, Mohammed},
  journal={IEEE access},
  volume={6},
  pages={52138--52160},
  year={2018},
  publisher={IEEE}
}

@article{younas2017non,
  title={Non-functional requirements elicitation guideline for agile methods},
  author={Younas, M and Jawawi, DNA and Ghani, I and Kazmi, R},
  journal={Journal of Telecommunication, Electronic and Computer Engineering (JTEC)},
  volume={9},
  number={3-4},
  pages={137--142},
  year={2017}
}

@inproceedings{alflen2020model,
  title={A Model for Evaluating Requirements Elicitation Techniques in Software Development Projects.},
  author={Alflen, Naiara C and Prado, Edmir PV and Grotta, Alexandre},
  booktitle={ICEIS (2)},
  pages={242--249},
  year={2020}
}

@article{pacheco2018requirements,
  title={Requirements elicitation techniques: a systematic literature review based on the maturity of the techniques},
  author={Pacheco, Carla and Garc{\'\i}a, Ivan and Reyes, Miryam},
  journal={IET Software},
  volume={12},
  number={4},
  pages={365--378},
  year={2018},
  publisher={Wiley Online Library}
}

@inproceedings{mishra2008successful,
  title={Successful requirement elicitation by combining requirement engineering techniques},
  author={Mishra, Deepti and Mishra, Alok and Yazici, Ali},
  booktitle={2008 First International Conference on the Applications of Digital Information and Web Technologies (ICADIWT)},
  pages={258--263},
  year={2008},
  organization={IEEE}
}

@article{hadar2014role,
  title={The role of domain knowledge in requirements elicitation via interviews: an exploratory study},
  author={Hadar, Irit and Soffer, Pnina and Kenzi, Keren},
  journal={Requirements Engineering},
  volume={19},
  pages={143--159},
  year={2014},
  publisher={Springer}
}

@inproceedings{speith2022XAITaxonomies,
author = {Speith, Timo},
title = {A Review of Taxonomies of Explainable Artificial Intelligence (XAI) Methods},
year = {2022},
isbn = {9781450393522},
publisher = {Association for Computing Machinery},
address = {New York, NY, USA},
doi = {10.1145/3531146.3534639},
booktitle = {Proceedings of the 2022 ACM Conference on Fairness, Accountability, and Transparency},
pages = {2239–2250},
numpages = {12},
location = {Seoul, Republic of Korea},
series = {FAccT '22}
}

@incollection{hypotheticalBiasPlott,
title = {Chapter 81 Experimental Evidence on the Existence of Hypothetical Bias in Value Elicitation Methods},
author = {Glenn W. Harrison and E. Elisabet Rutström},
editor = {Charles R. Plott and Vernon L. Smith},
booktitle = {Handbook of Experimental Economics Results},
publisher = {Elsevier},
address = "Amsterdam",
volume = {1},
pages = {752-767},
year = {2008},
issn = {1574-0722},
doi = {https://doi.org/10.1016/S1574-0722(07)00081-9}
}

@article{Deters2025quality,
title = {Exploring the means to measure explainability: Metrics, heuristics and questionnaires},
journal = {Information and Software Technology},
volume = {181},
pages = {107682},
year = {2025},
issn = {0950-5849},
doi = {https://doi.org/10.1016/j.infsof.2025.107682},
author = {Hannah Deters and Jakob Droste and Martin Obaidi and Kurt Schneider},
}

@article{Koehler2013,
    author     = {Koehler, Nicole and Vujovic, Olga and McMenamin, Christine},
    title      = {Healthcare Professionals' Use of Mobile Phones and the Internet in Clinical Practice},
    journal    = {Journal of Mobile Technology in Medicine},
    year       = {2013},
    volume     = {2},
    pages      = {3--13},
    month      = mar,
    issn       = {1839-7808},
    publisher  = {Deakin University},
    language   = {english},
    address    = {Washington, D.C.},
    note-comment       = {Open access}
}

@inproceedings{shklovski2014,
    author = {Shklovski, Irina and Mainwaring, Scott D. and Sk\'{u}lad\'{o}ttir, Halla Hrund and Borgthorsson, H\"{o}skuldur},
    title = {Leakiness and creepiness in app space: perceptions of privacy and mobile app use},
    year = {2014},
    isbn = {9781450324731},
    pages = {2347–2356},
    numpages = {10},
    booktitle = {CHI '14}
}

@Article{li2022,
AUTHOR = {Li, Zhouxiao and Koban, Konstantin Christoph and Schenck, Thilo Ludwig and Giunta, Riccardo Enzo and Li, Qingfeng and Sun, Yangbai},
TITLE = {Artificial Intelligence in Dermatology Image Analysis: Current Developments and Future Trends},
JOURNAL = {Journal of Clinical Medicine},
VOLUME = {11},
YEAR = {2022},
NUMBER = {22},
ARTICLE-NUMBER = {6826},
PubMedID = {36431301},
ISSN = {2077-0383},
}

@article{droste2025operationaltaxonomy,
  author    = {Jakob Droste and Hannah Deters and Martin Obaidi and Jil Klünder and Kurt Schneider},
  title     = {Framing What Can Be Explained -- An Operational Taxonomy for Explainability Needs},
  journal   = {Requirements Engineering},
  year      = {2025},
  issn      = {1432-010X}
}

@inproceedings{obaidi2025explainreqs,
  author    = {Martin Obaidi and Jannik Fischbach and Jakob Droste and Hannah Deters and Marc Herrmann and Jil Klünder and Steffen Krätzig and Hugo Villamizar and Kurt Schneider},
  title     = {Automatic Generation of Explainability Requirements and Software Explanations From User Reviews},
  booktitle = {2025 IEEE 33rd International Requirements Engineering Conference Workshops (REW)},
  year      = {2025},
  pages     = {49--58},
  doi       = {10.1109/REW66121.2025.00011}
}

@article{mavrepis2024,
    title        = {{XAI} for All: Can Large Language Models Simplify Explainable {AI}?},
    author       = {Mavrepis, Philip and Makridis, Georgios and Fatouros, Georgios and Koukos, Vasileios and Separdani, Maria Margarita and Kyriazis, Dimosthenis},
    year         = {2024},
    month        = jan,
    eprint       = {2401.13110},
    archiveprefix = {arXiv},
    primaryclass = {cs.AI},
    doi          = {10.48550/arXiv.2401.13110},
    journal      = {arXiv},
    volume       = {abs/2401.13110}
}

@article{stepin2021,
  author={Stepin, Ilia and Alonso, Jose M. and Catala, Alejandro and Pereira-Fariña, Martín},
  journal={IEEE Access}, 
  title={A Survey of Contrastive and Counterfactual Explanation Generation Methods for Explainable Artificial Intelligence}, 
  year={2021},
  volume={9},
  number={},
  pages={11974-12001},
}

@inproceedings{lubos2024,
author = {Lubos, Sebastian and Tran, Thi Ngoc Trang and Felfernig, Alexander and Polat Erdeniz, Seda and Le, Viet-Man},
title = {LLM-generated Explanations for Recommender Systems},
year = {2024},
isbn = {9798400704666},
booktitle = {UMAP Adjunct '24},
pages = {276–285},
numpages = {10},
}

@inproceedings{kabir2024,
    author     = {Kabir, Samia and Udo-Imeh, David N. and Kou, Bonan and Zhang, Tianyi},
    title      = {Is {Stack Overflow} Obsolete? An Empirical Study of the Characteristics of {ChatGPT} Answers to {Stack Overflow} Questions},
    booktitle  = {CHI '24},
    year       = {2024},
    articleno  = {935},
    pages      = {1--17},
    isbn       = {9798400703300}
}

@misc{ISOIECIEEE.2018,
  author       = {{ISO/IEC/IEEE}},
  title        = {Systems and Software Engineering --- Life Cycle Processes --- Requirements Engineering},
  year         = {2018},
  number       = {ISO/IEC/IEEE 29148:2018},
  type         = {Standard},
  publisher    = {{International Organization for Standardization}},
  address      = {Geneva},
  url          = {https://www.iso.org/standard/72026.html},
}

@article{Rupp.2009,
 author = {Rupp, Chris and Simon, Matthias and Hocker, Florian},
 year = {2009},
 title = {Requirements Engineering und Management},
 pages = {94--103},
 volume = {46},
 number = {3},
 issn = {1436-3011},
 journal = {HMD Praxis der Wirtschaftsinformatik},
 doi = {10.1007/BF03340367}
}

@article{Montgomery.2022,
 author = {Montgomery, Lloyd and Fucci, Davide and Bouraffa, Abir and Scholz, Lisa and Maalej, Walid},
 year = {2022},
 title = {Empirical research on requirements quality: a systematic mapping study},
 pages = {183--209},
 volume = {27},
 number = {2},
 issn = {0947-3602},
 journal = {Requirements Engineering},
 doi = {10.1007/s00766-021-00367-z}
}

@article{Heck.2018,
 author = {Heck, Petra and Zaidman, Andy},
 year = {2018},
 title = {A systematic literature review on quality criteria for agile requirements specifications},
 pages = {127--160},
 volume = {26},
 number = {1},
 issn = {0963-9314},
 journal = {Software Quality Journal},
 publisher = {Springer},
 doi = {10.1007/s11219-016-9336-4}
}

@book{Pohl.2005,
 author = {Pohl, Klaus and B{\"o}ckle, G{\"u}nter and {van der Linden}, Frank},
 year = {2005},
 title = {Software product line engineering: Foundations, principles, and techniques ; with 10 tables},
 address = {Berlin and Heidelberg},
 publisher = {Springer},
 isbn = {3-540-24372-0}
}

@inproceedings{withall2021readability,
  author    = {Withall, Amanda and Sagi, Eyal},
  title     = {The Impact of Readability on Trust in Information},
  year      = {2021},
  booktitle = {Proceedings of the 43rd Annual Meeting of the Cognitive Science Society},
  editor    = {Fitch, Tecumseh and Lamm, Claus and Leder, Helmut and Te{\ss}mar-Raible, Kristin},
  pages     = {2370--2376},
  volume     = {43},
  publisher = {Cognitive Science Society},
  address   = {Vienna, Austria},
}

@inproceedings{obaidi2025elicitation,
  author    = {Martin Obaidi and Jakob Droste and Hannah Deters and Marc Herrmann and Raymond Ochsner and Jil Klünder and Kurt Schneider},
  title     = {How to Elicit Explainability Requirements? A Comparison of Interviews, Focus Groups, and Surveys},
  booktitle = {2025 IEEE 33rd International Requirements Engineering Conference (RE)},
  year      = {2025},
  pages     = {167--178},
  doi       = {10.1109/RE63999.2025.00025}
}

@article{Norheim.2024,
 author = {Norheim, Johannes J. and Rebentisch, Eric and Xiao, Dekai and Draeger, Lorenz and Kerbrat, Alain and de Weck, Olivier L.},
 year = {2024},
 title = {Challenges in applying large language models to requirements engineering tasks},
 volume = {10},
 journal = {Design Science},
 doi = {10.1017/dsj.2024.8}
}

@article{Chen.2025,
  author  = {Chen, Banghao and Zhang, Zhaofeng and Langrené, Nicolas and Zhu, Shengxin},
  year    = {2025},
  title   = {Unleashing the potential of prompt engineering for large language models},
  pages   = {101260},
  volume  = {6},
  number  = {6},
  publisher = {Patterns (New York, N.Y.)},
  journal = {Patterns},
  doi     = {10.1016/j.patter.2025.101260 }
}

@incollection{Arora.2023,
  author    = {Arora, Chetan and Grundy, John and Abdelrazek, Mohamed},
  year      = {2024},
  title     = {Advancing Requirements Engineering through Generative AI: Assessing the Role of LLMs},
  editor    = {Nguyen-Duc, Anh and Abrahamsson, Pekka and Khomh, Foutse},
  booktitle = {Generative AI for Effective Software Development},
  publisher = {Springer},
  pages     = {129--148},
  doi       = {10.1007/978-3-031-55642-5_6}  
}

@inproceedings{deters2025explanationcatalog,
  author    = {Hannah Deters and Laura Reinhardt and Jakob Droste and Martin Obaidi and Kurt Schneider},
  title     = {Identifying Explanation Needs: Towards a Catalog of User-Based Indicators},
  booktitle = {RE},
  year      = {2025},
  pages     = {31--42},
  doi       = {10.1109/RE63999.2025.00014}
}

@inproceedings{obaidi2026usefulness,
  author    = {Martin Obaidi and Kushtrim Qengaj and Hannah Deters and Jakob Droste and Marc Herrmann and Kurt Schneider and Jil Klünder},
  title     = {Understanding Usefulness in Developer Explanations on Stack Overflow},
  booktitle = {Requirements Engineering: Foundation for Software Quality},
  year      = {2026},
  publisher = {Springer},
}

@inproceedings{droste2026immersive,
  author    = {Jakob Droste and Ronja Fuchs and Hannah Deters and Martin Obaidi and Alexander Dockhorn and Kurt Schneider},
  title     = {Immersive and Enjoyable Explanations - On Distinct Explainability Requirements in Games},
  booktitle = {Requirements Engineering: Foundation for Software Quality},
  year      = {2026},
  publisher = {Springer},
}

@inproceedings{droste2026misunderstandings,
  author    = {Jakob Droste and Hannah Deters and Carolin Kirchhoff and Lukas Nagel and Martin Obaidi and Kurt Schneider},
  title     = {Misunderstandings by Design: Using Erroneous Tutorials to Induce Mental Model Conflicts and the Need for Explanations},
  booktitle = {Requirements Engineering: Foundation for Software Quality},
  year      = {2026},
  publisher = {Springer},
}

@inproceedings{obaidi2025appfeatures,
  author    = {Martin Obaidi and Kushtrim Qengaj and Jakob Droste and Hannah Deters and Marc Herrmann and Jil Klünder and Elisa Schmid and Kurt Schneider},
  title     = {From App Features to Explanation Needs: Analyzing Correlations and Predictive Potential},
  booktitle = {2025 IEEE 33rd International Requirements Engineering Conference Workshops (REW)},
  year      = {2025},
  pages     = {99--106},
  doi       = {10.1109/REW66121.2025.00017}
}

@inproceedings{anders2023userfeedback,
  author    = {Michael Anders and Martin Obaidi and Alexander Specht and Barbara Paech},
  title     = {What Can Be Concluded from User Feedback? — An Empirical Study},
  booktitle = {2023 IEEE 31st International Requirements Engineering Conference Workshops (REW)},
  year      = {2023},
  pages     = {122--128},
  doi       = {10.1109/REW57809.2023.00027}
}

@inproceedings{anders2022mentalmodels,
  author    = {Michael Anders and Martin Obaidi and Barbara Paech and Kurt Schneider},
  editor    = {Vincenzo Gervasi and Andreas Vogelsang},
  title     = {A Study on the Mental Models of Users Concerning Existing Software},
  booktitle = {Requirements Engineering: Foundation for Software Quality},
  year      = {2022},
  publisher = {Springer International Publishing},
  address   = {Cham},
  pages     = {235--250},
  doi       = {10.1007/978-3-030-98464-9\_18}
}

@misc{obaidi2026multigolddataset,
  author       = {Obaidi, Martin},
  title        = {{Multidimensional Gold-Standard Dataset for Explanation Needs in App Reviews}},
  month        = may,
  year         = 2026,
  publisher    = {Zenodo},
  doi          = {10.5281/zenodo.20359756},
  url          = {https://doi.org/10.5281/zenodo.20359756}
}

@inproceedings{obaidi2026guidelineExp,
  author    = {Martin Obaidi and Jean-Carl Kremser and Hannah Deters and Jakob Droste and Marc Herrmann and Kurt Schneider},
  title     = {Writing Better Software Explanations: A Guideline-Based Approach},
  booktitle = {2026 IEEE 34th International Requirements Engineering Conference (RE)},
  year      = {2026},
  organization = {IEEE}
}

@misc{obaidi2026guidelineReqDataset,
  author       = {Obaidi, Martin and Droste, Jakob and Deters, Hannah and Herrmann, Marc and Krahl, Michel and Schneider, Kurt},
  title        = {{Replication Package: From Quality Properties to Practice: A Guideline and Workflow for Explainability Requirements}},
  month        = may,
  year         = 2026,
  publisher    = {Zenodo},
  doi          = {10.5281/zenodo.20396663},
  url          = {https://doi.org/10.5281/zenodo.20396663}
}

\end{document}